\documentclass[prl,floatfix,twocolumn,superscriptaddress]{revtex4}

\usepackage{graphicx}
\usepackage{bm}
\usepackage{amsmath}
\usepackage{url}

 \def\bk{{\bf k}} 

\def\bV{{\bf V}}  
  
 \def\bEDC{{\bf E}_{\mathrm{dc}}}

\def\frep{{f_\mathrm{rep}}}

\def\EDC{E_{\mathrm{dc}}}

\begin{document}
\title{Electric field-induced quantum interference control in a semiconductor:  \\ A new manifestation of the Franz-Keldysh effect}
\author{J.~K. Wahlstrand}
\altaffiliation[Current address: ]{Department of Physics, University of Maryland, College Park, MD 20742 USA}
\affiliation{JILA, National Institute of Standards and Technology, and the University of Colorado, Boulder, CO 80309 USA}
\author{H. Zhang}
\affiliation{JILA, National Institute of Standards and Technology, and the University of Colorado, Boulder, CO 80309 USA}
\affiliation{Department of Electrical, Computer, and Energy Engineering, University of Colorado, Boulder, CO 80309 USA}
\author{S. B. Choi}
\affiliation{JILA, National Institute of Standards and Technology, and the University of Colorado, Boulder, CO 80309 USA}
\author{S. Kannan}
\affiliation{JILA, National Institute of Standards and Technology, and the University of Colorado, Boulder, CO 80309 USA}
\affiliation{Department of Physics, University of Colorado, Boulder, CO 80309 USA}
\author{D. S. Dessau}
\affiliation{JILA, National Institute of Standards and Technology, and the University of Colorado, Boulder, CO 80309 USA}
\affiliation{Department of Physics, University of Colorado, Boulder, CO 80309 USA}
\author{J. E. Sipe}
\altaffiliation[Permanent address: ]{Department of Physics, University of Toronto, Toronto, Ontario, Canada M5S1A7}
\affiliation{JILA, National Institute of Standards and Technology, and the University of Colorado, Boulder, CO 80309 USA}
\author{S. T. Cundiff}
\affiliation{JILA, National Institute of Standards and Technology, and the University of Colorado, Boulder, CO 80309 USA}
\affiliation{Department of Electrical, Computer, and Energy Engineering, University of Colorado, Boulder, CO 80309 USA}
\affiliation{Department of Physics, University of Colorado, Boulder, CO 80309 USA}
\date{\today}

\begin{abstract}
In (100)-oriented GaAs illuminated at normal incidence by a laser and its second harmonic, interference between one- and two-photon absorption results in ballistic current injection, but not modulation of the overall carrier injection rate.
Results from a pump-probe experiment on a transversely biased sample show that a constant electric field enables coherent control of the carrier injection rate.
We ascribe this to the nonlinear optical Franz-Keldysh effect and calculate it for a two-band parabolic model.
The mechanism is relevant to centrosymmetric semiconductors as well.
\end{abstract}

\maketitle

Enabled by the accelerating development of ultrafast lasers, coherent control using interference between quantum pathways \cite{shapiro_coherent_2003} has become an increasingly important tool for the optical control of matter.
Experiments using multiphoton pathways have included directional ionization in atoms \cite{yin_asymmetric_1992}, photodissociation \cite{sheehy_phase_1995} and alignment \cite{de_field-free_2009} in molecules, and injection of ballistic charge currents \cite{dupont_phase-controlled_1995,hache_observation_1997} and pure spin currents \cite{stevens_quantum_2003} in semiconductors, to list just a few examples.
The symmetry of the system to be controlled constrains the processes that can occur, and an important but largely neglected extension is the use of constant or nearly constant external fields to change a coherent control process from forbidden to allowed.
The simplest such field, and likely the most useful for potential applications in the solid state, is an electric field.
While it has been shown that a constant (dc) electric field enables the control of the ionization rate of an atom by a two-color process \cite{manakov_dc_1999,gunawardena_atomic_2007}, the use of such a simple control parameter has not yet been exploited in other systems.
Here we show that a dc field can enable new coherent control processes in semiconductors.

Interference between one- and two-photon pathways in semiconductors leads to the injection of a ballistic charge or spin current, the direction of which is related to the polarization of the light and the relative phase between superimposed optical beams with frequencies $\omega$ and $2\omega$.
Referred to as $1+2$ quantum interference control (QUIC), this process has been used to create a carrier-envelope phase sensitive photodetector \cite{fortier_carrier-envelope_2004} and to study subpicosecond current dynamics \cite{kerachian_dynamics_2007,zhao_dynamics_2008}.
It is also possible to control the carrier \emph{population} via $1+2$ QUIC \cite{fraser_quantum_1999}, in a process arising from the imaginary part of the second-order nonlinear susceptibility $\chi^{(2)}$.
This process may be observed in a zincblende semiconductor such as GaAs when the optical fields are aligned along certain crystal axes.
One could easily imagine that, just as in electric field-induced second harmonic generation \cite{lee_nonlinear_1967}, a dc field could combine with the always present third-order susceptibility $\chi^{(3)}$ to produce an effective $\chi^{(2)}$ that enables $1+2$ population control.
Here, we present experimental evidence that such a process does in fact occur, and we develop a simple theory relating it to the Franz-Keldysh effect \cite{franz__1958,keldysh__1958,aspnes_electric_1967}, a modification of the optical properties of a semiconductor due to the acceleration of photoexcited carriers.

\begin{figure}
\includegraphics[width=8.5cm]{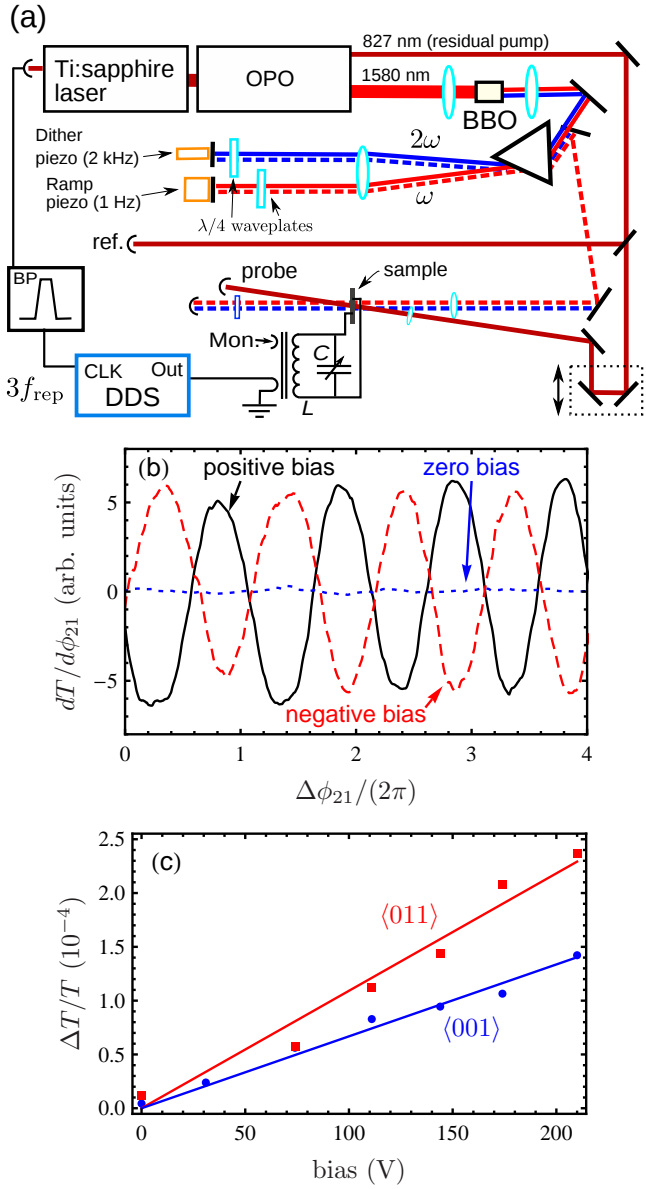}
\caption{(color online) Measurement of field-induced QUIC.  (a)  Schematic of the experimental apparatus, including the two-color interferometer and the radio frequency biasing scheme.
OPO: optical parametric oscillator, BBO: $\beta$-barium borate crystal, BP: band pass filter, DDS: direct digital synthesis board.
(b) The modulation of the transmitted probe power at the dither frequency as the phase parameter $\phi_{21} = \phi_{2\omega}-2\phi_{\omega}$ is swept by the slow ramp piezo.
The measured signal for $\bEDC \parallel \langle 001 \rangle$ is shown for a positive bias (solid black), negative bias (dashed red), and zero bias (blue dotted).
(c) The amplitude of the differential transmission signal is shown as a function of the bias voltage for $\bEDC \parallel \langle 001 \rangle$ (blue dots) and $\bEDC \parallel \langle 011 \rangle$ (red squares).
The lines are linear fits to the data points.
}
\label{expt}
\end{figure}

We performed an all-optical measurement on (100) GaAs samples at room temperature.
A diagram of the experimental apparatus and the results are shown in Fig.~\ref{expt}.
The technique is similar to what was used to observe conventional population control in (111) GaAs \cite{fraser_quantum_1999}, with the addition of a bias across the sample.
A two-color pump pulse injects carriers through one- and two-photon absorption and QUIC.
The transmission of a probe pulse depends on the photo-excited carrier density \cite{lee_room-temperature_1986}, and we look for changes that depend on the phase parameter $\phi_{21} = \phi_{2\omega}-2\phi_{\omega}$.
The light source for the pump beam is an optical parametric oscillator (OPO) using cesium titanyl arsenate (CTA), synchronously pumped by a mode-locked Ti:sapphire laser with repetition rate $\frep = 76$ MHz.
The signal output of the OPO, centered at 1580 nm, is doubled using a $\beta$-barium borate (BBO) crystal, and the two harmonics are split by a prism into two paths, the relative delay of which can be adjusted using mirrors mounted on piezoelectric transducers.
The harmonics are recombined using the same prism. 
We use the residual pump light, centered at 827 nm, as a probe beam.
The pulse widths of the OPO fundamental, second harmonic, and probe pulses are approximately 160 fs, 220 fs, and 120 fs, respectively.

We use a 1 $\mu$m thick, undoped GaAs epilayer grown by molecular beam epitaxy.
The GaAs sample is attached to the sapphire disk using transparent epoxy, and the substrate is then removed by chemical etching.
A 100 nm thick insulating SiO$_2$ layer is deposited on the surface.
Using photolithography, Au electrodes separated by 100 $\mu$m are then patterned onto the sample.
To apply an effective dc electric field $\EDC$, we use a radio frequency (rf) bias synchronized to $\frep$.
This technique results in an electric field in the plane of the sample \cite{wahlstrand_uniform-field_2010} while avoiding highly nonuniform fields encountered when carriers are injected from electrodes into semi-insulating samples \cite{ralph_trap-enhanced_1991}.
We derive a harmonic of $\frep$ from a photodiode and use it as a clock for direct digital synthesis (DDS) of a synchronized, variable phase bias waveform at $\frep$.
A resonant $LC$ circuit passively enhances the voltage across the electrodes.
The phase of the waveform is optimized so that the peak of the waveform occurs when the optical pulse arrives at the sample.
The sign of the field may be reversed by changing the phase of the rf signal by $\pi$ using the DDS.
Linear electroabsorption studies on the sample studied here, at the same bias voltage and optical power, show that we achieve a field strength of approximately 20 kV/cm.
Two samples were studied: one with the rf field pointing along a $\langle 001 \rangle$ crystal direction and the other with the field along $\langle 011 \rangle$.

The pump and probe beams are focused on a common spot between the electrodes.
The $\omega$ beam average power is 160 mW and the $2\omega$ beam average power is 6 mW.
Part of the probe beam is split off before the sample and used as a reference for balanced detection.
The spot size of the pump beams is roughly 25 $\mu$m, and the spot size of the probe beam half that.
To detect only the contribution from QUIC processes, we dither $\phi_{21}$ using a mirror mounted on a piezoelectric transducer.
A lock-in amplifier referenced to the dither frequency measures only the changes in the probe transmission $T$ due to changes in $\phi_{21}$; the signal is proportional to $dT/d\phi_{21}$.
Using another mirror mounted on a piezoelectric transducer, we ramp the phase parameter over 0.5 s and average the data over approximately 100 ramp cycles.
Results are shown in Fig.~\ref{expt}b with a positive and a negative bias, as well as with no bias.
For light at $\omega$ and $2\omega$ normally incident on the (100) GaAs sample, for which population control enabled by the crystal symmetry is forbidden \cite{fraser_quantum_1999}, we observe modulation of the transmitted probe beam \emph{only} when the bias is applied.

The magnitude of the bias can be controlled by adjusting the output amplitude of the DDS.
Results are shown in Fig.~\ref{expt}c.
The signal is consistent with a linear dependence on bias, for both positive and negative bias (the latter is not shown).
The dependence on the polarization of the optical fields was also studied by rotating the waveplates.
Because of the polarization dependent reflection at the interfaces of the prisms in the interferometer, it is not feasible to continuously adjust the polarization.
However, in a study of a few different polarization configurations we found that the signal drops by more than a factor of 4 when either or both of the $\omega$ or $2\omega$ beams are polarized perpendicular to the dc field direction.
As described later, this result is consistent with the theory.

\begin{figure}
\includegraphics[width=8cm]{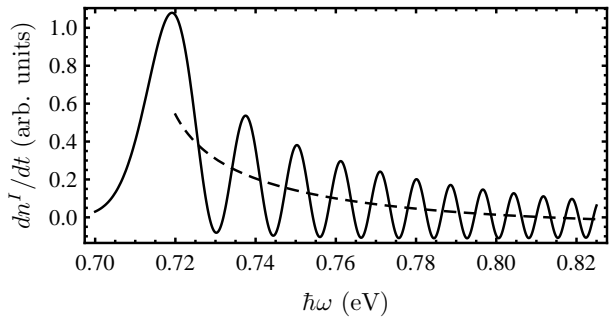}
\caption{Calculation of field-induced $1+2$ quantum interference control of carrier injection in GaAs in the parabolic band approximation, with no lifetime broadening (solid line) and in the limit of large broadening (dashed line).  The optical fields are linearly polarized parallel to the constant field $\EDC=20$ kV/cm.}
\label{theory}
\end{figure}

For one-photon absorption, the change in optical properties of a bulk semiconductor is typically dominated by the Franz-Keldysh effect (FKE) \cite{franz__1958,keldysh__1958,aspnes_electric_1967}.
To calculate QUIC in the presence of a static electric field, we extend a theory of the one-photon FKE \cite{wahlstrand_theory_2010} to multiphoton processes.
In the limit of long optical pulses, one can derive a Fermi Golden Rule expression for the rate of carrier injection due to QUIC.
This is of the form
$\dot{n}^{\mathrm{I}} = \eta^{jlm} (\omega) E^i_{2\omega} E^l_\omega E^m_\omega e^{i(\phi_{2\omega}-2\phi_{\omega})}+c.c.$, where
$\eta^{jlm} (\omega)$ is a tensor that describes the efficiency of the process as a function of $\omega$.
For two parabolic bands separated by a direct band gap $\hbar \omega_g$, $\eta^{jlm} (\omega)$ can be found analytically.
We assume a $\bk$-independent interband velocity matrix element $\bV_{cv}$.
The conduction band and valence band effective masses are $m_c$ and $m_v$, respectively, and the reduced effective mass $\mu = m_c m_v/(m_c+m_v)$.
For a constant field pointing along $\hat{\mathbf{z}}$, we find
\begin{multline}
\eta^{zzz} (\omega) = \frac{e^3 \EDC^2}{2\hbar^3\omega^4\Omega} |V^z_{cv}|^2 \left(\frac{\mathrm{Ai}^2 (-\frac{2\omega-\omega_g}{\Omega})}{2\Omega} \right. \\ \left.+ \frac{-\frac{2\omega-\omega_g}{\Omega} \mathrm{Ai}^2 (-\frac{2\omega-\omega_g}{\Omega})-[\mathrm{Ai}'(-\frac{2\omega-\omega_g}{\Omega})]^2}{\omega} \right),
\label{lowtemp}
\end{multline}
where $\mathrm{Ai}(x)$ is the Airy function and $\Omega=(e^2 \EDC^2/2\mu\hbar)^{1/3}$ is the electro-optic frequency.
This parabolic band approximation (PBA) result is shown in Fig.~\ref{theory} for the parameters of GaAs.
Because the tensor is real, the injection rate is proportional to $\cos(\phi_{2\omega}-2\phi_{\omega})$.

As with the one-photon absorption spectrum \cite{aspnes_electric_1967}, the spectrum of the field-induced QUIC injection tensor displays Franz-Keldysh oscillations.
The theory assumes no decoherence or lifetime broadening; we expect that damping tends to wash out the oscillations, as it does for one-photon absorption \cite{aspnes_electric_1967}.
Using the asymptotic form of the Airy function for large $\omega$, which gives the spectrum in terms of $\sin^2(x)$ and $\cos^2(x)$, and replacing them with the average value 0.5, we find
\begin{multline}
\eta^{zzz} (\omega) = \\ \frac{\sqrt{2} e^4  \mu^{1/2}}{2\pi\hbar^{9/2} \omega^4 } |V^z_{cv}|^2 \left( \frac{1}{4\sqrt{2\omega-\omega_g}} - \frac{\sqrt{2\omega-\omega_g}}{ \omega} \right) \EDC
\label{qic_w_damping}
\end{multline}
for $2\omega>\omega_g$.
This theoretical expression, more relevant than Eq.~(\ref{lowtemp}) for a sample at room temperature, is plotted as a dashed line in Fig.~\ref{theory}.
We find that the injection rate due to QUIC changes sign at $(4/7)\hbar \omega_g$ in the PBA.
Using analytical expressions for the one-photon and two-photon absorption in the PBA, neglecting field-induced changes in those (which are of order $\EDC^2$), one can show that 
\begin{equation}
\frac{\dot{n}^I}{\dot{n}} = \frac{6e^2 \EDC E_\omega^2 E_{2\omega} \omega \left(4\omega_g-7\omega\right)}{16e^2 E_\omega^4 (2\omega-\omega_g)^2+3E_{2\omega}^2 \mu \hbar \omega^4 (2\omega-\omega_g)}
\label{percent}
\end{equation}
for $2\omega>\omega_g$.

A realistic model for the band structure displays a dependence on the direction of the fields with respect to the crystal axes, as well as more subtle effects due to nonparabolicity \cite{wahlstrand_theory_2010}.
A preliminary calculation using a 14-band $\mathbf{k}\cdot\mathbf{p}$ model has the same qualitative spectral shape as the PBA result, with the zero in the injection rate occurring at a lower energy.
Since the process we describe arises from a fourth-rank tensor (to lowest order in the DC field), the polarization dependence can be predicted knowing the zincblende crystal symmetry.
The nonzero components are $\chi_{xxxx}$ (all fields parallel), $\chi_{xxyy}$, and $\chi_{xyxy}$.
The largest effect is expected for all fields parallel, as observed in the experiment.


The modulation of the pump-probe differential transmission due to QUIC in the $\langle 001 \rangle$ sample for the probe arriving 300 fs after the pump is approximately 2.0\%.
The expected carrier density modulation from Eq.~(\ref{percent}), using the fluences of the $\omega$ and $2\omega$ pulses in the experiment, is 0.6\%.
In an optically thin sample, the fractional change in the pump-probe transmission would equal the fractional change in the carrier density.
Because the thickness of our sample is comparable to the coherence length between the $\omega$ and $2\omega$ beams ($\approx$1.2 $\mu$m \cite{fraser_quantum_2003}), the amplitude of the modulation varies as a function of depth, complicating the analysis.
There are other complicating factors:
The QUIC signal decays for longer pump-probe time delays.
We attribute this to the relaxation of a non-uniform distribution of carriers due to the combination of dispersion, multiple reflections, and nonuniform absorption for the $2\omega$ pulse.
A cascaded process \cite{stevens_enhanced_2005}, arising in this case from electric field-induced second harmonic generation (EFISH) \cite{lee_nonlinear_1967} followed by optical interference, also contributes to the signal.
Considering these experimental issues and the simplifying assumptions made in the theory, the agreement is reasonable.
We have observed additional evidence of field-induced QUIC in experiments on low-temperature grown GaAs and Er-doped GaAs that use electrical \cite{hache_observation_1997,wahlstrand_electrical_2006} rather than optical detection.
Biasing the electrodes used to read out the injected photocurrent results in an enhancement in the signal that is inconsistent with current injection alone.

In summary, we have shown that a constant electric field enables control of the carrier injection rate by interference of one- and two-photon absorption in a semiconductor, in a new manifestation of the Franz-Keldysh effect.
While the use of slowly-varying electric fields along with optical pulses has not been much explored in coherent control, the results here indicate that the experiments are feasible and the results for absorption in a semiconductor are interpretable in terms of an extension of Franz-Keldysh theory.
We note that similar processes have been predicted \cite{manakov_dc_1999} and observed \cite{gunawardena_atomic_2007,bolovinos_one-_2008} in atomic systems.
The process described here is very different, because it relies on the acceleration of the photo-excited carriers (Franz-Keldysh effect) rather than the perturbation of bound states (Stark effect) by the electric field.
While the experiment here showed a linear dependence of the coherent control efficiency on the dc field strength, experiments at low temperature or for energies near the half-band gap should show a highly nonlinear response characteristic of the nonperturbative nature of the Franz-Keldysh effect.

Because it results in exotic carrier distributions such as ballistic charge and spin currents, coherent control via QUIC is a promising strategy for studying transport across metal-semiconductor interfaces, of great importance in improving electronic devices.
Since DC fields are often encountered in such structures, understanding their effect on QUIC is essential in realizing the potential of this diagnostic tool.
In this paper we have studied the electro-optic coherent control of carrier population, perhaps the simplest solid state coherent control process, but also one of the weakest:
even for crystals and polarization schemes where coherent control of population is allowed, it is small.
But coherent control processes such as those for charge and spin current injection are more robust effects, and this work suggests that their modulation by DC fields are promising directions for future studies, from both fundamental and technological perspectives.

We thank R.~Smith, J.~Pipis, and R.~Snider for assistance in the early stages of the experiment, R.~Mirin for providing the GaAs sample, and T.~Reber for technical assistance with the OPO.
S.T.C.~is a staff member in the NIST Quantum Physics division.


\end{document}